\documentclass[aps,prb,twocolumn,showpacs,groupedaddress]{revtex4}
\usepackage{graphicx}
\usepackage{amssymb}
\usepackage{bm}

\begin{document}

\vspace*{-2.0cm}
{\tt PHYSICAL REVIEW B {\bf 68}, 174414 (2003)}
\vspace{0.5cm}

\title{Magnetic relaxation and dipole-coupling-induced magnetization\\
in nanostructured thin films during growth: A cluster Monte Carlo study}
\author{R. Brinzanik}
\email[electronic address: ]{brinzani@physik.fu-berlin.de}
\author{P.J. Jensen}
\author{K.H. Bennemann}
\affiliation{Institut f{\"u}r Theoretische Physik, Freie Universit{\"a}t Berlin, 
Arnimallee 14, D-14195 Berlin, Germany}

\date{\today}

\begin{abstract}
For growing inhomogeneous thin films with an island nanostructure 
similar as observed in experiment, we determine the nonequilibrium 
and equilibrium remanent magnetization. The single-island magnetic 
anisotropy, the dipole coupling, and the exchange interaction  between
magnetic islands are taken into account within a micromagnetic model.  
A cluster Monte Carlo method is developed which includes coherent 
magnetization changes of connected islands. This causes a 
fast relaxation towards equilibrium for irregularly connected systems. 
We analyse the transition from dipole coupled islands at low coverages to a 
strongly connected ferromagnetic film at high coverages during film growth. 
For coverages below the percolation threshold, the dipole interaction 
induces a collective magnetic order with ordering temperatures of 
1 -- 10~K for the assumed model parameters.  
Anisotropy causes blocking temperatures of 10 --
100~K and thus pronounced nonequilibrium effects. 
The dipole coupling leads to a somewhat slower magnetic relaxation. 
\end{abstract}    


\maketitle

\section{\label{intro}Introduction}
The investigation of low-dimensional magnetic nanostructures 
has become a very active field of current research.\cite{himpsel} 
The controlled preparation of different nanostructures 
allows for the investigation of 
a variety of interesting magnetic properties.\cite{sun,shen97b,bode00}
The dependence of these properties on the nanostructure and 
magnetic interactions in such systems is still not well understood. 
So far, no consistent theoretical analysis has been
performed for the magnetic behavior of an ultrathin film during growth, 
ranging from an island-type structure to a smooth film, in particular 
the influence of structural disorder on the magnetic properties. 
In this study, we present Monte Carlo (MC) calculations of the 
nonequilibrium and equilibrium magnetization of growing inhomogeneous 
films. Different coverages below and above the percolation threshold and 
different magnetic interactions are taken into account. 
This analysis needs a newly developed cluster MC method
including coherent rotations of neighboring magnetic islands. 

To illustrate the problem, we discuss for instance an ultrathin Co film 
grown on a Cu(001) substrate.\cite{bovensiepen,poulopoulos} 
This system shows a perfect layer-by-layer growth for coverages 
$\Theta > 2$~monolayers (ML). Here, the exchange 
coupling results in a large Curie temperature 
$T_\mathrm{C}\gtrsim 350$~K. Below the measured percolation 
coverage of $\Theta_\mathrm{P} = 1.7$~ML, randomly positioned 
Co islands with a considerable admixture of Cu atoms 
are observed, exhibiting a comparably strong remanent magnetization 
up to temperatures as large as $\sim150$~K. 
At $\Theta_\mathrm{P}$, a jump of $T_\mathrm{C}$ 
of about 100~K occurs.\cite{bovensiepen} The important question arises 
which mechanism causes the strong remanence of an island-type film 
for coverages $\Theta < \Theta_\mathrm{P}$. On the one hand, this could be 
induced by nonequilibrium blocking effects due to single-island 
anisotropies which impede magnetic relaxation.
On the other hand, an equilibrium magnetization can originate from long-range 
magnetic interactions. We like to discuss these two limiting cases. 

An ensemble of magnetically isolated single-domain 
islands behaves like a superparamagnet.
The single-island anisotropy causes a finite time-dependent 
magnetization below the nonequilibrium blocking temperature $T_\mathrm{b}$.
For thin Co/Cu(001) films this temperature is
estimated to be  $T_\mathrm{b}\approx5$~K,\footnote{We have 
assumed the measured in-plane anisotropy 
$K=0.01$~meV/atom,\cite{heinrich} a typical number of atoms $N=1000$ 
per Co island at $\Theta<\Theta_\mathrm{P}$,\cite{poulopoulos} 
a measurement time $\tau_\mathrm{m}=100$~sec, and the attempt frequency
$\Gamma_o=10^9$~sec$^{-1}$;\cite{neel,brown1} $k_\mathrm{B}$ is the 
Boltzmann constant.} 
following from the Arrhenius-N{\'e}el ansatz  
$k_\mathrm{B}\,T_\mathrm{b}=NK/\ln(\tau_\mathrm{m}\,\Gamma_o)$.
\cite{neel,brown1} Note that an island size dispersion will influence 
the relaxational behavior and $T_\mathrm{b}$ sensitively. 
Second, a finite magnetization may also originate from a collectively 
ordered state in thermal  equilibrium. Such a magnetic state, not 
necessarily a collinear one, of an ensemble of isolated islands 
results from long-range magnetic interactions.\cite{cowburn}  
We consider here the magnetic dipole coupling between islands.
\footnote{In addition to the magnetic dipole 
coupling other long-range interactions can be present such as the 
indirect exchange (RKKY-) interaction in case of 
metallic substrates, and the superexchange for insulating substrates.}
The corresponding ordering temperature $T_\mathrm{C}$ should be comparable to 
the average dipole energy per island which is, however, difficult to determine 
for an irregular system.  
Assuming two neighboring disk-shaped Co islands containing $N$ atoms
each, the dipole energy $E_\mathrm{dip}$ of this island pair is 
proportional to $\sqrt{N}$. For  $N=1000$, we estimate 
$E_\mathrm{dip}\approx6$~K.\footnote{Consider 
two disk-shaped Co islands with $N$ atoms each. 
Assume that the distance $R$ between their centers is comparable to 
their diameters, $R\approx D\propto a_o\,\sqrt{N}$. Then the point-dipole 
energy of this island pair is 
$E_\mathrm{dip}\approx (N\,\mu_\mathrm{at})^2/R^3 \approx 
0.623\;(\mu_\mathrm{at}^2/a_o^3)\,\sqrt{N}\approx6$~K, 
with $\mu_\mathrm{at}= 2.0\,\mu_\mathrm{B}$ the 
atomic magnetic moment measured for a 2-ML Co film,\cite{srivastava} 
$\mu_\mathrm{B}$ the Bohr magneton, $a_o=2.5$~{\AA}{} the Co interatomic 
distance, and $N=1000$.} 

Note that both simple estimates for the temperatures $T_\mathrm{b}$ and 
$T_\mathrm{C}$ disagree with experimental observations. Hence, 
improved calculations are needed 
taking better into account the inhomogeneous film structure characterized by 
varying island sizes, shapes, and positions. In particular for
coverages close to $\Theta_\mathrm{P}$, the island coagulation 
leads to a larger effective island size $N_\mathrm{eff}$ and thus 
to a larger average dipole energy 
$E_\mathrm{dip}\propto \sqrt{N_\mathrm{eff}}$. 
Correspondingly, also the blocking temperature 
$T_\mathrm{b}\propto N_\mathrm{eff}K\,$ 
due to anisotropy will be larger than the single-island estimate 
$T_\mathrm{b}\approx5$~K. Thus, for a disordered film structure one may expect 
a remanent magnetization at much higher temperatures than obtained by
these simple estimates. 

For the investigation of the magnetic relaxation, the magnetic anistropy 
and the dipole interaction are taken into account. In case of coagulated 
islands, the exchange coupling between islands has to be considered as well. 
For three-dimensional systems of interacting magnetic particles, 
either longer\cite{dormann} as well as shorter
\cite{morup}
relaxation times as compared to noninteracting ensembles have been calculated. 
Experimentally, longer relaxation times for increasing interparticle 
interactions have been measured.\cite{mamiya,poddar} The existence of a 
collective, spin-glass-like magnetic ordering was discussed.\cite{jonsson}
For  two-dimensional systems only few investigations have been
reported, also indicating longer relaxation times for increasing 
interaction strengths.\cite{lottis,poddar}

In this paper, we report on MC calculations of the nonequilibrium and 
equilibrium remanent magnetization for nanostructured thin films during growth
as function of coverage, temperature, and MC time.\cite{chantrell,dompub,phd} 
By application of a modified Ising model, which allows one to take into 
account magnetization dynamics, the blocking as well as the 
ordering temperatures are determined. Of particular concern is the 
consideration of structurally inhomogeneous systems ranging from 
isolated islands to smooth ferromagnetic films. Numerical 
simulations are unavoidable since the low symmetry of these systems and
the complicated nature of the involved magnetic interactions 
preclude analytical approaches.  
However, the application of the common MC technique runs into a severe 
problem. Assigning a ``super'' spin to every island magnetic moment 
(Stoner-Wohlfarth model\cite{stonerw47}), conventional 
single-spin-flip algorithms yield an extremely slow and unrealistic relaxation 
towards equilibrium for coverages where the islands are partly coagulated.
\footnote{Consider for instance the magnetic rotation 
$\uparrow\uparrow\;\rightharpoonup\;\downarrow\downarrow$ of an 
isolated island cluster consisting of two connected islands. Assuming 
Ising-like magnetic moments, inevitably the single-spin-flip method 
applies a \textit{subsequent} rotation of the single island spins 
through the intermediate state $\uparrow\downarrow$. However, such a process 
is very unlikely due to the large increase of exchange energy of 
this intermediate state. Obviously, a proper treatment of magnetic relaxation 
requires the inclusion of a \textit{coherent} rotation of the island
pair. This is performed within a cluster-spin-flip algorithm 
which takes into account such simultaneous spin rotations of 
connected islands.}
Hence, we apply a \textit{cluster-spin-flip algorithm,} 
which includes simultaneous rotations of magnetic moments of connected 
islands.\cite{wolff,davis} We emphasize that 
with this method the relaxation behavior and the equilibrium magnetization 
are calculated efficiently. For all film coverages, 
the cluster-spin-flip method enables an appropriate analysis of the 
influence of the anisotropy and dipole coupling  
besides the dominating exchange interaction. 

The film growth, the micromagnetic 
model for the calculation of the magnetic properties, and the 
cluster MC method are described in Sec.~\ref{model}. In 
Sec.~\ref{results}, we test the cluster algorithm for the obtained 
island nanostructure and present results for the remanent
magnetization, as well as for the blocking and ordering temperatures.  
A conclusion is given in Sec.~\ref{conclusion}. 

\section{\label{model}Model and MC simulation}
\subsection{\label{eden}Growth mode and micromagnetic model}
For the simulation of the island-type growing film, 
we use the simple solid-on-solid Eden model.\cite{landaubinder,dompub} 
Within this model, each additional atom is deposited on an island
perimeter site $i$ with probability 
$p(q_i,z_i)   \propto \exp[A(z_i)\,\sqrt{q_i}]$,  
where $q_i$ is the local coordination number and $z_i$ the layer index. 
A bilayer island  growth mode is assumed yielding an island 
structure similar as observed  for epitaxial Co/Cu(001).\cite{poulopoulos}
A $2 \times 500 \times 500)$ fcc-(001) unit cell with lateral 
periodic boundary conditions is applied. The island density 
is $\rho=0.0025$~islands per site, resulting in $Z=625$ 
randomly distributed islands in the unit cell. For the 
ratio of binding parameters,\cite{dompub} we use $A(1)/A(2)=0.989$. 

For the obtained atomic structures, a micromagnetic model for the 
total (free) energy of a system of interacting magnetic islands 
is applied:\cite{chantrell,dompub} 
\begin{eqnarray} \label{e2}
\lefteqn{E = -\frac{1}{2} \sum_{i>j}L_{ij}(\Theta)\;
             \gamma_{ij}(\Theta,T)\;\bm{S}_i \bm{S}_j {}}\nonumber\\ 
    & & {} + \sum_{i>j}
         \frac{\mu_i(\Theta,T) \,\mu_j(\Theta,T)}{r_{ij}^5}
         \left[ r_{ij}^2 \;\bm{S}_i \bm{S}_j 
   -3\,( \bm{r}_{ij} \bm{S}_i)\,(\bm{r}_{ij} \bm{S}_j) \right] \nonumber\\  
    & & {} - \sum_i N_i(\Theta)\;K_i(\Theta,T)\;(S_i^x)^2 {} \;, 
\end{eqnarray}  
with $\Theta$ being the film coverage and $T$ the temperature.
Each magnetic island with $N_i(\Theta)$ atoms is treated as a 
Stoner-Wohlfarth particle\cite{stonerw47} with a single giant 
magnetic moment $\mu_i=\mu_\mathrm{at}\,m_i\, N_i$, whose direction is 
confined to the film plane, where $\mu_\mathrm{at}$ is the atomic magnetic 
moment. The unit vector $\bm{S}_i=\bm{\mu}_i/\mu_i$ 
characterizes the magnetic moment direction of the $i$th island. 
The first term in Eq.~(\ref{e2}) represents the 
magnetic domain wall energy between connected islands, 
with $L_{ij}$ the number of bonds between islands $i$ and $j$, and 
$\gamma_{ij}$ the domain wall energy per atomic bond. The second 
term is the long-range magnetic
dipole interaction between the island magnetic moments $\mu_i$ where 
$r_{ij}=|\bm{r}_i-\bm{r}_j|$ is the distance between the centers 
of islands $i$ and $j$. The point-dipole energy is calculated by 
applying the Ewald summation technique over all periodically arranged 
unit cells of the thin film.\cite{jensen97,phd} The last term 
denotes the uniaxial in-plane 
anisotropy energy with $K_i$ the anisotropy per atomic spin. 
Due to this anisotropy we allow for only two
stable directions for each island moment ($S_i^x=\pm1$). 
Thus, our system refers to a \textit{modified Ising model}, for which 
during magnetization reversal a possible anisotropy energy 
barrier is taken into account, hence allowing the consideration of 
magnetization dynamics. Finally, due to the finite 
exchange coupling $J$ between neighboring atomic spins, 
the \textit{internal island magnetization} $m_i(\Theta,T)$ 
is taken into account within a mean-field approximation. 
This leads to temperature-dependent effective anisotropy coefficients
$K_i(\Theta,T)$ and domain wall energy densities
$\gamma_{ij}(\Theta,T)$, as described in greater detail in Ref.~17.  

Equation~(\ref{e2}) describes a system of dipole-coupled single islands at 
low film coverages $\Theta\ll\Theta_\mathrm{P}$ as well as 
a connected ferromagnetic film at high coverages $\Theta\gg\Theta_\mathrm{P}$. 
We point out that the transition between these extremal cases 
during the film growth is described within the same  model. 
The assumption of individual magnetic islands with 
varying interactions is a good approximation as long as the system is 
laterally nanostructured, whereas for smooth films (here 
$\Theta \approx 2.0$~ML) it 
represents an unphysical discretization of the system. 

\subsection{\label{algorithm}Cluster MC method}
The magnetic equilibrium and nonequilibrium 
properties are calculated by performing kinetic cluster MC simulations.
Especially, close to the percolation coverage $\Theta_\mathrm{P}$, 
most of the magnetic islands are connected to neighboring islands and form  
large but still finite clusters. A single-spin-flip (SSF)
algorithm\cite{landaubinder} for such an irregular atomic structure 
yields a very slow relaxation towards thermodynamic equilibrium, since 
\textit{subsequent} flips of island magnetic moments in this cluster  
of connected islands, as considered by SSF updates, 
are strongly hindered by the exchange energy, see Ref.~34. 
Thus, a rotation of the entire island cluster is very unlikely, and
its dependence on dipole interaction and anisotropy 
is strongly underestimated. 
For an improved simulation of the magnetic relaxation, a \textit{coherent} 
or \textit{simultaneous} rotation of the spins in these clusters 
has to be taken into account. For this purpose, we propose a 
cluster-spin-flip (CSF) algorithm\cite{wolff,davis} in the present study. 
 
In a first step of each MC update, a cluster $\mathcal{C}_\nu$ consisting of $\nu$ 
connected islands is constructed by the following scheme:
 
(a) Choose randomly a single island $i$, representing the first (smallest) 
island cluster $\mathcal{C}_1=\{i\}$. 

(b) Add a random second island $j$ 
which is connected to island $i$ ($L_{ij}\not=0)$, 
forming the second island cluster $\mathcal{C}_2=\{i,j\}$.

(c) Construct subsequently larger island clusters $\mathcal{C}_\nu$ by 
adding a randomly chosen island to the preceding cluster 
$\mathcal{C}_{\nu-1}$, provided that this island is connected to at 
least one of the $\nu-1$ islands of $\mathcal{C}_{\nu-1}$. 

(d) Continue this construction procedure till either 
no additional adjacent islands are present or if a maximum allowed number 
$\lambda_\mathrm{max}$ of islands in the cluster is reached. 
From this procedure, we obtain a set of $\lambda\leq \lambda_\mathrm{max}$ 
island clusters $\{\mathcal{C}_1,\ldots,\mathcal{C}_\lambda\}$.  

(e) Out of this set choose one cluster $\mathcal{C}_\nu$ 
with weight $\omega_\nu$ for probing ($\sum_\nu \omega_\nu=1$). 

A Monte Carlo step (MCS) is defined by the usual condition that 
$Z$ islands in the system 
are probed. Employing 
a cluster $\mathcal{C}_\nu$ containing $\nu$ islands considers 
the portion $\nu/Z$ of the system in a single update. To ensure 
that probing large clusters does not dominate the relaxation process, 
we assign the weight $\omega_\nu=1/\nu$ for choosing $\mathcal{C}_\nu$ 
out of the set $\{\mathcal{C}_1,\ldots,\mathcal{C}_\lambda\}$. 
This definition implies that within 
a single MCS no additional relaxation channels 
are opened by the consideration of island cluster flips. 
\footnote{The described CSF algorithm satisfies the condition 
of detailed balance. This is guaranteed by 
the fact that the probabilities for construction and choice of
the cluster $\mathcal{C}_\nu$ are the same for both flip directions, 
and by the used flip rates which already obey  
detailed balance. Ergodicity is maintained since any spin state 
can be reached due to the allowance of single-spin flips.}
We emphasize that not only is the largest possible island cluster 
$\mathcal{C}_\lambda$ probed for flipping, but 
all island clusters out of the corresponding set are considered. 
The island moments within an island cluster need not to be parallel. 

In the second step of each update, all $\nu$ island spins of the chosen 
cluster $\mathcal{C}_\nu$ are probed for a coherent flip. 
The corresponding flip rate $\Gamma_\nu$ 
is calculated in the usual way as if these $\nu$ connected islands 
form a single large island.\cite{dompub}   
From Eq.~(\ref{e2}), the magnetic energy of this island cluster 
as function of the in-plane angle $\phi$ is given by 
\begin{equation}
\epsilon_\nu(\phi) = \frac{E_\nu(\phi)}{\mathcal{K}_\nu} =
-2 h_\nu \cos \phi - \cos^2 \phi \;, \label{e7}
\end{equation}
with the reduced magnetic field 
\begin{equation}
h_\nu = \frac{\sum_{kl}S_k^x S_l^x \big[L_{kl}\,\gamma_{kl} + 
  \mu_k \mu_l\left( 4x_{kl}^2-2y_{kl}^2 \right)/r_{kl}^5\big]}
{4 \,\mathcal{K}_\nu}  \label{e8} 
\end{equation}
and the total anisotropy energy of the cluster
$\mathcal{K}_\nu=\sum_{k=1}^\nu N_k K_k$.
Here, the $k$ sum runs over all spins \textit{inside}, and the $l$ sum
over all spins \textit{outside} the island cluster $\mathcal{C}_\nu$. 
We have neglected the dipole sums $\sum_{kl}(x_{kl}\,y_{kl})/r_{kl}^5$ 
which are usually smaller than the sums
$\sum_{kl}(x_{kl})^2/r_{kl}^5$ and $\sum_{kl}(y_{kl})^2/r_{kl}^5$.  

The Ising-like states $S_i^x=\pm 1$ of $\mathcal{C}_\nu$ 
represent either energy minima which are separated by an anisotropy 
energy barrier, 
or refer to an energy maximum and minimum. The respective  energy 
barriers for the forward and backward transitions are given by 
\begin{eqnarray} 
\Delta E_\nu^{(1)}=(h_\nu\pm1)^2 \, \mathcal{K}_\nu &\quad&
\mathrm{for} \quad |h_\nu| < 1 \;, \label{e9} \\ 
\Delta E_\nu^{(2)}=\pm \,4 \,h_\nu \, \mathcal{K}_\nu &\quad&
\mathrm{for} \quad |h_\nu| \geq 1 \;. \label{e11} 
\end{eqnarray} 
The flip rate 
$\Gamma_\nu^{(1)}$ of the island spin cluster $\mathcal{C}_\nu$    
to overcome $\Delta E_\nu^{(1)}$ is calculated from the common 
Arrhenius-N{\'e}el ansatz.\cite{neel,brown1} 
We use a constant prefactor $\Gamma_o=10^9$~sec$^{-1}$ which determines 
the time unit of the magnetic relaxation in kinetic MC simulations. 
\footnote{A constant attempt frequency $\Gamma_o$ is widely applied in 
literature. The justification of this approach is given in 
Refs.~7 and 8 where $\Gamma_o$ is calculated using different 
approaches. $\Gamma_o$ is found to vary weakly with temperature and 
local fields. Note that the applied transition rates depend only weakly 
on the exact value of $\Gamma_o$.} 
The latter case is treated with the usual Metropolis-type 
rate, using the same prefactor.\cite{landaubinder}

The growing thin film is characterized by 
a large amount of nonequivalent lattice sites, corresponding to 
a large number of different interaction parameters. 
Since little is known about these values, we use in our simulation 
averaged quantities for the magnetic parameters which are fixed as
follows, using as an example the Co/Cu(001) thin-film system.
The atomic magnetic moments are set 
to $\mu_\mathrm{at}=2.0\,\mu_\mathrm{B}$.\cite{srivastava}
The domain wall energy $\gamma$ is adjusted to give the 
observed Curie temperature of the ferromagnetic {\em long-range\/} order
of $T_\mathrm{C}=355$~K of a 2-ML Co/Cu(001) film,\cite{bovensiepen} 
yielding $\gamma=5.6$~meV/bond. The exchange interaction for the 
calculation of the internal island magnetic ordering is set equal to 
$J=7.0$~meV/bond.\cite{dompub} For the uniaxial anisotropy, two different 
values $K=0.1$ and 0.01~meV/atom are investigated. 

In this study, we determine the remanent magnetization 
$M_\mathrm{rem}(\Theta,T,t)=\sum_iN_i\,S_i^x(t)\,m_i(\Theta,T)/\sum_i N_i$
of the growing thin film, where $t$ is the MC time in units of MCS. 
The simulation starts from a completely aligned island spin state. 
The choice of this initial state refers to experiments which 
saturate the magnetic system by an external magnetic field and 
determine the remanent magnetization after removal 
of the field.\cite{shen97b,bovensiepen}
We have no evidence that magnetic arrangements end up in metastable 
states during relaxation when starting from a saturated state.
In addition, we calculate the equilibrium magnetization 
$M_\mathrm{eq}(\Theta,T)$, which is obtained by averaging 
$M_\mathrm{rem}(\Theta,T,t)$ over a range of 500~MCS after the 
system has become equilibrated. $M_\mathrm{rem}(\Theta,T,t)$ and 
$M_\mathrm{eq}(\Theta,T)$ are averaged over at least 20 different 
structural runs.  The magnetizations are 
given in units of a saturated monolayer (i.e., $\Theta=1$~ML) at $T=0$. 
Since the finite-sized unit cell undergoes eventually 
total magnetic reversals during MC probing, accidental 
cancellation of a finite $M_\mathrm{rem}$ and $M_\mathrm{eq}$ 
during structural and temporal averaging may occur. To avoid this, we 
use in this study merely the \textit{absolute values} $|M_\mathrm{rem}|$ 
and $|M_\mathrm{eq}|$.

\section{\label{results}Results and discussion}
\subsection{Island growth and CSF algorithm}
Snapshots of the atomic structure during thin film growth, resulting from 
our growth model, are shown in Ref.~17.
The resulting static atomic structure is similar to the one observed 
for the Co/Cu(001) system.\cite{poulopoulos} 
In the initial stages of growth, randomly located islands
with almost rectangular shapes are obtained. With increasing film coverages, 
the single islands start to coagulate and form  
island clusters with a still finite size. By analysing the
percolation probability using the Hoshen-Kopelman algorithm, 
we yield a percolation coverage of about 
$\Theta_\mathrm{P} \approx 0.9$~ML.\cite{stauffer,hoshen} 
Continued film growth leads to a connected thin film.  
In this coverage range, the system still exhibits a distinct 
irregular nanostructure. Isolated island clusters vanish 
rapidly upon further adatom deposition. 
The coverage $\Theta = 2.0$~ML corresponds to 
a smooth magnetic film with two closed layers. 
 
\begin{figure*}[h]
\includegraphics[bb=95 198 520 780,width=9cm,clip]{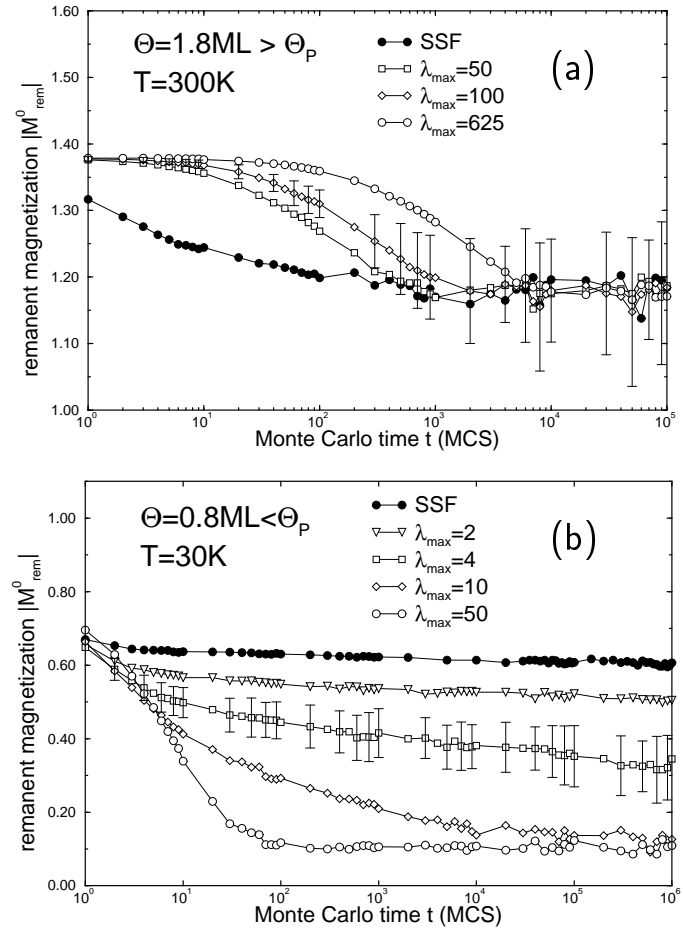}
\caption{\label{fig1}Comparison of the calculated film magnetization
using the single-spin-flip (SSF) and cluster-spin-flip (CSF) algorithm.
Only the exchange coupling between magnetic islands is considered. 
The remanent magnetization $|M_\mathrm{rem}^0|$ as function of 
MC time $t$ for two film coverages (a) $\Theta=1.8$~ML and (b) 
$\Theta=0.8$~ML above and below the percolation coverage 
$\Theta_\mathrm{P}\approx0.9$~ML is shown. Within the CSF method, 
different maximum numbers $\lambda_\mathrm{max}$ of coherently
flipping island magnetic moments are used.}
\end{figure*}

At first, we investigate the effect of the cluster-spin-flip MC method 
on the simulation of the remanent magnetization 
$|M_\mathrm{rem}(\Theta,T,t)|$. Here we consider only the exchange 
interaction. For a strongly connected film 
($\Theta_\mathrm{P} \ll \Theta=1.8$~ML we test whether by use of CSF 
$|M_\mathrm{rem}(\Theta,T,t)|$ relaxes into the correct equilibrium 
value $|M_\mathrm{eq}(\Theta,T)|$, starting from a fully aligned state. 
As can be seen from Fig.~\ref{fig1}(a), this condition is fulfilled, since 
different maximum allowed numbers $\lambda_\mathrm{max}$ of islands 
in the cluster lead to the same equilibrium value 
as for the single-spin-flip MC method. 
The larger the number $\lambda_\mathrm{max}$ is, the slower is
the relaxation. This property is caused 
by the fact that in this coverage and temperature range 
the magnetic relaxation is mainly provided by flips of single islands or
small island clusters. As mentioned in Sec.~\ref{algorithm}, no additional 
relaxation channels are opened by use of CSF, hence 
the number of single-spin-flip attempts becomes reduced in favor of unprobable 
cluster-spin-flip ones.
Examples for the error bars are also given. The statistical
error, which is similar for all forthcoming figures, 
results mainly from averaging over different structural realizations 
of the unit cell and could be reduced by using larger unit cells.  

We point out that the main improvement of the CSF with respect to the SSF 
method is obtained for coverages $\Theta\lesssim\Theta_\mathrm{P}$, characterized by 
a considerable amount of island cluster formation, and which 
is very difficult to be studied analytically. 
In Fig.~\ref{fig1}(b), the magnetic relaxation for $\Theta=0.8$~ML 
is depicted using different $\lambda_\mathrm{max}$. 
The equilibrium magnetization $M_\mathrm{eq}^0(\Theta,T)$ 
should vanish for $\Theta<\Theta_\mathrm{P}$, 
since long-range magnetic interactions are neglected here. 
Due to the use of the absolute value a finite but small 
$|M_\mathrm{eq}^0(\Theta,T)|$ is obtained in our calculations.
The SSF algorithm exhibits an extremely slow
magnetic relaxation toward $|M_\mathrm{eq}^0|$. Even after $10^6$~MCS the 
remanent magnetization is relaxed only to $|M_\mathrm{rem}^0|=0.60$. 
The reason is that this method considers very unfavorable 
intermediate states. Already the allowance of a few coherently flipping 
island spins results in a much faster relaxation. The relaxational behavior 
converges rapidly with increasing  $\lambda_\mathrm{max}$. 
Using the CSF with $\lambda_\mathrm{max}=50$ or larger 
the equilibrium is reached already after $\sim 100$~MCS. 
To obtain a fast equilibration, the closer the coverage 
to $\Theta_\mathrm{P}$  the larger the value chosen for 
$\lambda_\mathrm{max}$.    
The CSF algorithm leads to a much faster equilibration also for coverages
$\Theta\gtrsim\Theta_\mathrm{P}$. In this coverage range, island 
clusters are still present which have only weak links to other clusters. 
In the following investigations, 
we set $\lambda_\mathrm{max}$ equal to the number of single islands
$Z=625$, except for coverages $\Theta\gg\Theta_\mathrm{P}$ where we yield
a better performance for $\lambda_\mathrm{max}=100$.

Hence, by use of the SSF algorithm the exchange interaction 
grossly dominates the MC simulations for strongly inhomogeneous systems. 
This is avoided by applying CSF, allowing thus for the investigation 
of the effect of the much weaker anisotropy and dipole interaction. 

\subsection{Effect of interactions}
\begin{figure*}[h]
\includegraphics[bb=101 198 514 780,width=9cm,clip]{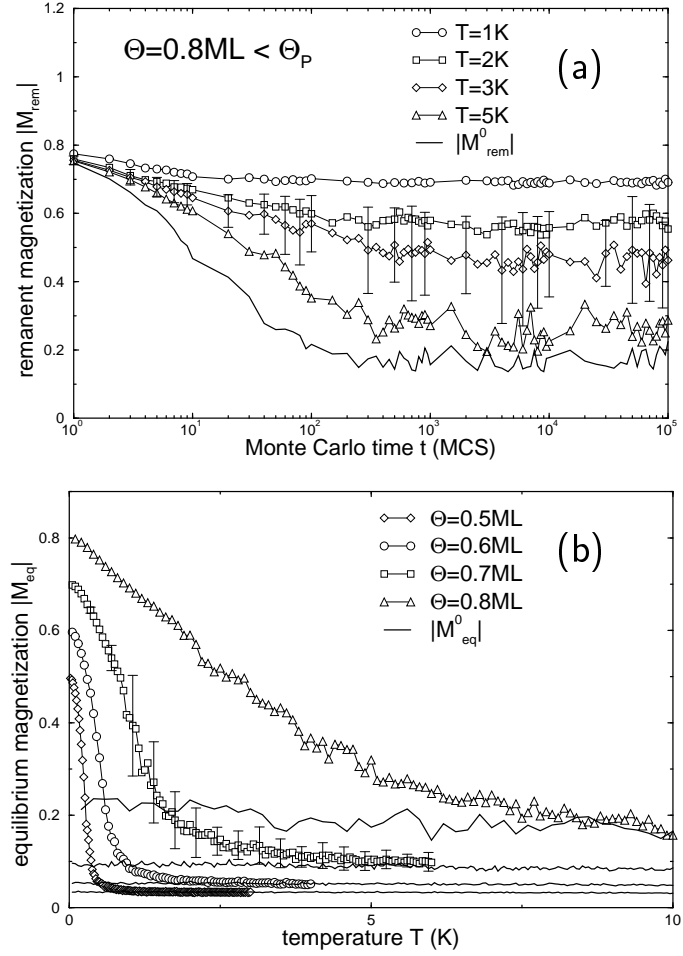}
\caption{\label{fig2}Long-range magnetic ordering due to dipole interaction
for coverages $\Theta$ below the percolation coverage $\Theta_\mathrm{P}$.
Only dipole and exchange interactions are included. The quantities 
$|M_\mathrm{rem}^0|$ and $|M_\mathrm{eq}^0|$ neglect the dipole interaction.  
(a) Remanent magnetization $|M_\mathrm{rem}|$ as function of MC time $t$ 
for $\Theta=0.8$~ML $<\Theta_\mathrm{P}$ and different temperatures $T$. 
(b) Equilibrium magnetization $|M_\mathrm{eq}|$ as function of temperature 
$T$ for different coverages $\Theta<\Theta_\mathrm{P}$.}
\end{figure*}

First, we study the combined effect of the dipole and the exchange
interaction on the film 
magnetization for coverages $\Theta<\Theta_\mathrm{P}$.
We determine equilibrium properties which within our model are not
influenced by the anisotropy.  
In Fig.~\ref{fig2}(a), we present 
$|M_\mathrm{rem}(\Theta,T,t)|$ for different temperatures $T$ as a
function of MC time $t$. The coverage is assumed to be $\Theta=0.8$~ML. 
Starting from the fully aligned state 
$|M_\mathrm{rem}(\Theta,T,t=0)|=0.8$, the remanent film magnetization 
relaxes fast to its equilibrium value. 
For the assumed temperatures the dipole coupling leads to a net magnetization 
$|M_\mathrm{rem}| > |M_\mathrm{rem}^0|$ where for $|M_\mathrm{rem}^0|$ 
the dipole interaction is neglected. After several hundred MCS 
equilibration is obtained for the dipole-coupling induced $|M_\mathrm{rem}|$ 
which then stays stable within the simulation time. 
We emphasize that it is impossible 
to obtain these and the following results with conventional SSF algorithms.

In Fig.~\ref{fig2}(b), the equilibrium magnetization $|M_\mathrm{eq}(\Theta,T)|$
is shown as function of temperature $T$ for different coverages $\Theta$.
For low temperatures, clearly a magnetic ordering 
due to the dipole interaction is seen. 
The larger the coverage is, the larger is
the ordering effect, since with an increasing $\Theta$
the average island cluster size and thus the average dipole coupling energy
increases. Above the \textit{ordering temperatures} 
$T_\mathrm{C}(\Theta)$, the magnetizations 
$|M_\mathrm{eq}(\Theta,T)|$ reach the corresponding values 
$|M_\mathrm{eq}^0(\Theta,T)|$ as calculated without the dipole interaction.
Due to the use of absolute values, $|M_\mathrm{eq}^0(\Theta,T)|$ 
stays always finite. $T_\mathrm{C}(\Theta)$ is estimated by  
extrapolating the linear part of $|M_\mathrm{eq}|$ to $|M_\mathrm{eq}^0|$.
The ordering temperature even for the largest
investigated coverage is quite small, yielding 
$T_\mathrm{C}\approx6$~K for $\Theta=0.8$~ML.
The rounding of $|M_\mathrm{eq}|$  near the ordering
temperature is caused by (i) the finite unit cell size, (ii)
the use of the absolute value, (iii) the presence of 
island size and -position dispersions, and (iv) the average 
over 20 different realizations of the unit cell. 

The existence of a long-range magnetic ordering due to the
dipole interaction has been calculated for \textit{periodic} lattices. 
\cite{macisaac00}
In our study, we find that also within an \textit{irregular} island 
system below the percolation threshold the dipole interaction 
leads to a magnetic ordering, indicated by a 
net magnetization $|M_\mathrm{eq}|>|M_\mathrm{eq}^0|$.   
Such a collective state is expected to be 
spin-glass-like, as discussed for three-dimensional magnetic particle 
systems,\cite{jonsson} and which needs a further 
investigation. We remark that the obtained ordering temperatures will 
increase by considering a noncollinear island
magnetization beyond $S_i^x=\pm1$,\cite{jensen02a} 
by taking into account the finite island extension for the dipole interaction 
beyond the point-dipole approximation,\cite{jensen02b} or by consideration 
of densely packed three-dimensional particles.

\begin{figure*}[h]
\includegraphics[bb=184 240 431 782,width=8cm,clip]{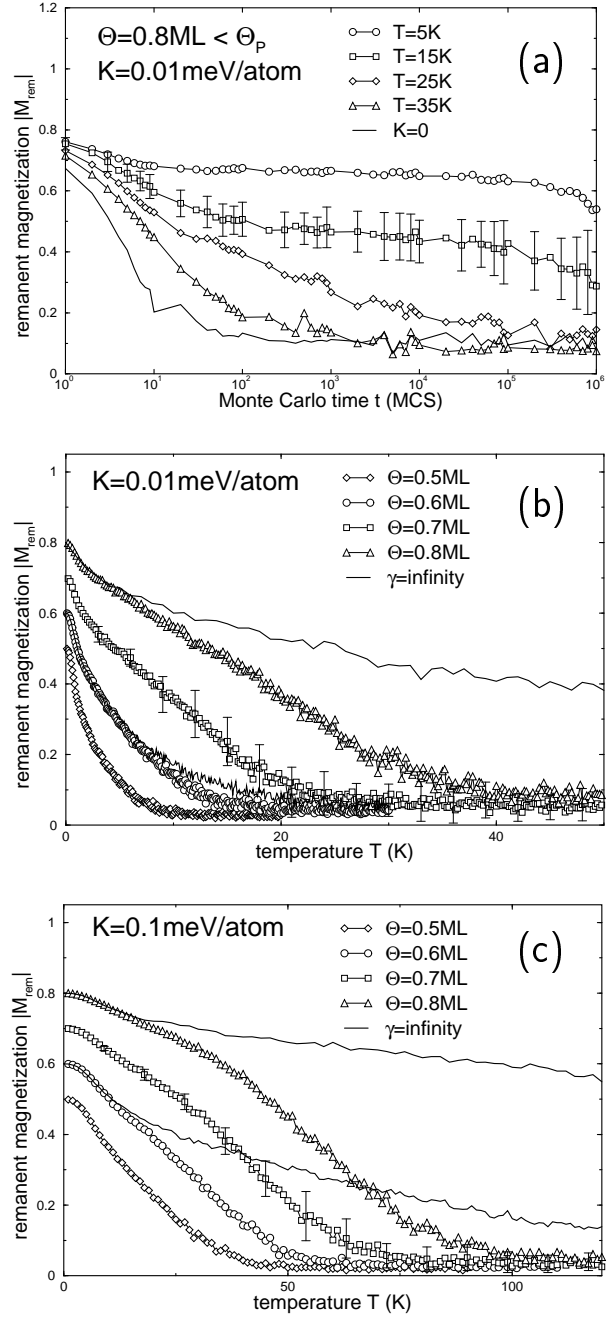}
\caption{\label{fig3}Relaxation of the remanent magnetization for 
coverages $\Theta$ below the percolation coverage $\Theta_\mathrm{P}$. 
Only anisotropy and exchange interaction are taken into account. 
(a) Remanent magnetization $|M_\mathrm{rem}|$ as function of 
MC time $t$ for $\Theta=0.8$~ML $<\Theta_\mathrm{P}$, anisotropy 
$K=0.01$~meV/atom, and different temperatures $T$. The full line refers 
to $K=0$ and $T=35$~K. Furthermore, (b) and (c) show  
the remanent magnetization $|M_\mathrm{rem}|$ after $t=1000$~MCS as function 
of temperature $T$ for $K=0.01$ and 0.1~meV/atom. The full lines are 
calculated with an infinite domain wall energy $\gamma$.} 
\end{figure*}

In Fig.~\ref{fig3}, we investigate the influence of the magnetic anisotropy 
and the exchange interaction on the magnetic relaxation for coverages 
$\Theta < \Theta_\mathrm{P}$. Here, the dipole coupling is neglected. 
In Fig.~\ref{fig3}(a), the remanent magnetization 
$|M_\mathrm{rem}(\Theta,T,t)|$ is given as function of MC time $t$
for different temperatures $T$ and for $K=0.01$~meV/atom. 
The coverage is assumed to be $\Theta=0.8$~ML. 
Starting from the fully aligned state, at first 
$|M_\mathrm{rem}(\Theta,T,t)|$ drops rapidly due to relaxation 
of single islands and small island clusters. The further relaxation happens 
much more slowly since here larger island clusters have to be reversed. 
For $T\gtrsim25$~K, the magnetization $|M_\mathrm{rem}(\Theta,T,t)|$ 
reaches within the depicted time range the curve 
$|M_\mathrm{rem}^0|$ as calculated for $K=0$.  
 
In Figs.~\ref{fig3}(b) and \ref{fig3}(c), we show $|M_\mathrm{rem}(\Theta,T,t)|$ 
after $t=1000$~MCS for different coverages as function of temperature, using 
the anisotropy parameters $K=0.01$ and $0.1$~meV/atom. 
With increasing temperature, the magnetization $|M_\mathrm{rem}(\Theta,T,t)|$ 
approaches  the equilibrium value $|M_\mathrm{eq}^0|$. The corresponding 
\textit{blocking temperatures} $T_\mathrm{b}(\Theta,K)$ are obtained 
by extrapolating the linear part of $|M_\mathrm{rem}|$ to $|M_\mathrm{eq}^0|$. 
A rounding of $|M_\mathrm{rem}|$ near $T_\mathrm{b}$ is observed due to the
same reasons as discussed in connection with Fig.~\ref{fig2}.  

We emphasize that for a connected island structure 
an increase of the anisotropy $K$ by a factor of 10 does not necessarily 
lead to an increase of $T_\mathrm{b}(\Theta,K)$ by the same factor
as obtained from the Stoner-Wohlfarth model.\cite{stonerw47} 
This is caused by \textit{internal cluster excitations}, 
i.e., creation or motion of domain walls inside island clusters. 
To discuss this, we have performed additional calculations 
for an infinite domain wall energy $\gamma$, indicated by the 
full lines in Figs.~\ref{fig3}(b) and \ref{fig3}(c), hence allowing only for coherent 
island cluster rotations. Above a certain temperature, the curves for finite 
and infinite $\gamma$ deviate, 
since then internal cluster excitations become effective. 
For $K=0.01$~meV/atom and coverage $\Theta=0.6$~ML,  the difference 
between these curves is small, thus the magnetic relaxation happens mainly 
via coherent rotation. In contrast, for 
$\Theta=0.8$~ML or for the larger anisotropy $K=0.1$~meV/atom, obviously 
both relaxation processes are present. 

Which relaxation process is effective at a given temperature 
is determined by its energy barrier $\Delta E$. 
For a coherent rotation of an isolated island cluster, $\Delta E$ is given by 
its total anisotropy energy $\mathcal{K}_\nu$.  In contrast, 
$\Delta E$ for an internal cluster excitation consists of 
both the anisotropy of the actually reversed islands and 
the domain wall energy, see Eqs.(\ref{e9}) and (\ref{e11}). 
By closer investigation, we found that a
particular relaxation process becomes effective above a temperature amounting 
to 5 -- 10 \% of $\Delta E$. In the temperature ranges  
$T<40$~K for $K=0.01$~meV/atom and $T<100$~K for $K=0.1$~meV/atom,  
each internal cluster excitation consists mainly of reversing  
only one or two islands. For markedly larger temperatures, 
the internal cluster 
excitations will become more complex, depending in a complicated 
way on the nanostructure. 

\begin{figure*}[h]
\includegraphics[bb=94 79 540 725,width=6.5cm,angle=-90,clip]{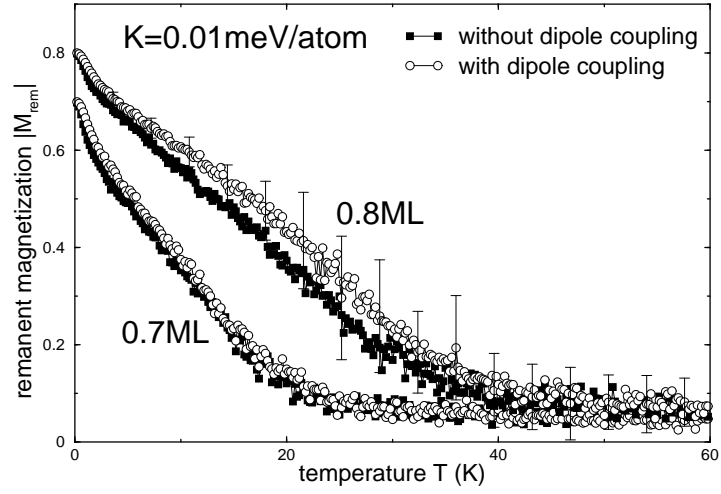}
\caption{\label{fig4}Influence of magnetic anisotropy and dipole coupling on
the remanent magnetization $|M_\mathrm{rem}(T)|$ as shown in 
Fig.~\ref{fig3}(b). Results for two coverages $\Theta$ below the 
percolation coverage $\Theta_\mathrm{P}$ are presented. }
\end{figure*}

The influence of the dipole coupling on the
relaxation behavior is discussed in Fig.~\ref{fig4}. 
For $\Theta\lesssim\Theta_\mathrm{P}$ and $K=0.01$~meV/atom,
the dipole interaction results in a small increase of $|M_\mathrm{rem}|$
and $T_\mathrm{b}$. Thus, the dipole interaction leads to a {\em slower\/} 
magnetic relaxation. Interestingly, this effect is visible in the whole 
temperature range up to $T_\mathrm{b}$, and is not limited to those small 
temperatures where the dipole coupling induces a magnetic ordering, 
see Fig.~\ref{fig2}. The increase of $|M_\mathrm{rem}|$ 
will become larger if a stronger dipole coupling is assumed, for 
example for larger island magnetic moments. 
Experimentally, a similar effect was observed for two-dimensional arrays of 
interacting magnetic nanoparticles with random anisotropy axes.\cite{poddar}
A more detailed investigation of this property is needed. 

\begin{figure*}[h]
\includegraphics[bb=94 79 540 722,width=6.5cm,angle=-90,clip]{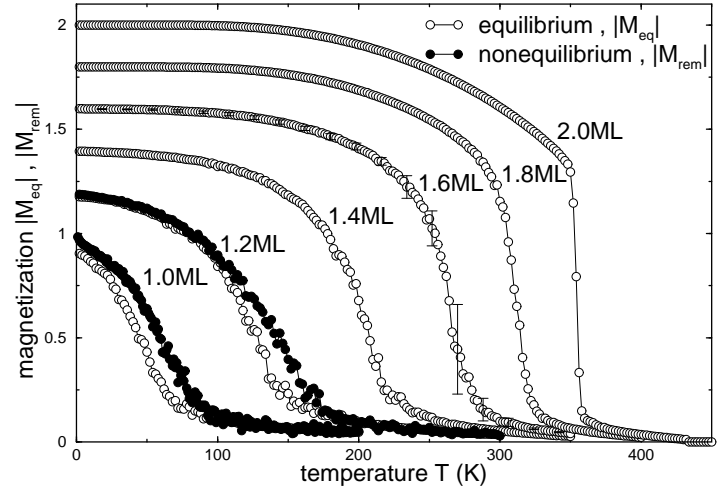}
\caption{\label{fig5} Long-range ferromagnetic ordering due to exchange 
interaction for different coverages $\Theta$ above the percolation 
coverage $\Theta_\mathrm{P}$. The equilibrium magnetization 
$|M_\mathrm{eq}|$ is shown as function of temperature $T$. For comparison, 
for $\Theta=1.0$ and 1.2~ML also the remanent magnetization 
$|M_\mathrm{rem}|$ after $t=1000$~MCS is depicted for an anisotropy 
$K=0.01$~meV/atom.} 
\end{figure*}

Next, we investigate the magnetization for 
coverages above the percolation threshold, $\Theta>\Theta_\mathrm{P}$.
In Fig.~\ref{fig5}, the equilibrium magnetization $|M_\mathrm{eq}(\Theta,T)|$ 
is shown as function of temperature for different coverages. 
Here, the exchange coupling causes a fast magnetic relaxation and 
a strong ferromagnetic long-range order. For large coverages and at low 
temperatures, the behavior of 
$|M_\mathrm{eq}|$ is reigned by the decrease of the 
internal island magnetization $m_i(\Theta,T)$, whereas 
at elevated temperatures a strong decay of $|M_\mathrm{eq}|$ is caused by 
the disturbance of the island spin alignment.  
The resulting ordering temperatures $T_\mathrm{C}(\Theta)$ are deduced 
from the inflection points of $|M_\mathrm{eq}(\Theta,T)|$. 
In addition, for $\Theta=1.0$~ML and $1.2$~ML we show the nonequilibrium
remanent magnetization $|M_\mathrm{rem}|$ after $t=1000$~MCS, 
considering $K=0.01$~eV/atom. The corresponding
blocking temperatures $T_\mathrm{b}(\Theta)$ are markedly larger than 
$T_\mathrm{C}(\Theta)$ 
in the coverage range $\Theta\gtrsim \Theta_\mathrm{P}$, where the 
nanostructure of the percolated thin film is still very irregular 
and nonequilibrium effects due to anisotropy barriers are pronounced. 
For films with larger coverages, having a higher connectivity 
between islands, the exchange coupling results in a fast magnetic 
relaxation and thus the temperature difference between $T_\mathrm{b}$ 
and $T_\mathrm{C}$ is small. A very weak ordering effect due to the 
dipole interaction is visible only for very low temperatures 
and coverages $\Theta\gtrsim\Theta_\mathrm{P}$. Here, still a few 
isolated islands or island clusters 
exist which are coupled to the percolating cluster by the dipole interaction. 

\begin{figure*}[h]
\includegraphics[bb=84 70 540 722,width=6.5cm,angle=-90,clip]{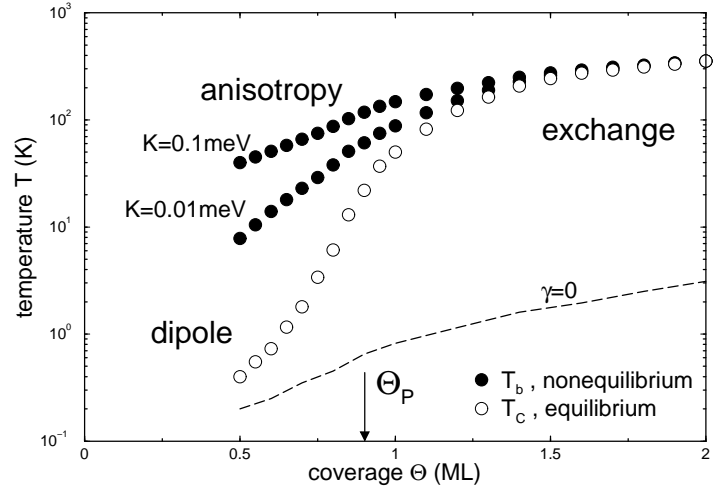}
\caption{\label{fig6}Semi-logarithmic plot of the magnetic ordering 
temperature $T_\mathrm{C}$ and the blocking temperature $T_\mathrm{b}$ 
as functions of the film coverage $\Theta$. The temperatures  
$T_\mathrm{C}$ and  $T_\mathrm{b}$ are extracted from the preceding figures. 
The entire investigated coverage range below and above the percolation
coverage $\Theta_\mathrm{P}$ is shown. The different magnetic interactions 
dominate in different coverage and temperature ranges as indicated. 
The dashed curve refers to the dipole-coupling-induced ordering 
temperature neglecting the domain wall energy $\gamma$ between islands.}
\end{figure*}

In Fig.~\ref{fig6}, we summarize the most important results of this study.
The (nonequilibrium) blocking temperature $T_\mathrm{b}(\Theta,K)$
and the (equilibrium) ordering temperature $T_\mathrm{C}(\Theta)$ are 
presented as functions of coverage $\Theta$ in the whole investigated 
growth range.
For a better visualization, a logarithmic temperature scale is applied.
$T_\mathrm{b}$ is determined for two different anisotropies $K=0.01$ and 
$0.1$~meV/atom and $t=1000$~MCS. Below the percolation coverage 
$\Theta_\mathrm{P}$, the dipole interaction induces small ordering 
temperatures $T_\mathrm{C}$ of the order of 1 -- 10~K for the assumed 
model parameters. Due to the coagulation of islands with increasing coverage 
the exchange interaction becomes more important, 
since it couples single islands to magnetically aligned large clusters. 
This results in a strong increase of $T_\mathrm{C}$ 
in particular close to $\Theta_\mathrm{P}$. 
This behavior has been observed in experiments on Co/Cu(001) ultrathin films
(``$T_\mathrm{C}$-jump'').\cite{bovensiepen,poulopoulos} 
For percolated thin films, the ordering temperature
is of the order of $100$ -- $300$~K and is, within the accuracy of our 
calculations, exclusively determined by the exchange coupling. The slope 
of $T_\mathrm{C}(\Theta)$ for $\Theta > \Theta_\mathrm{P}$ is not as steep 
as for $\Theta < \Theta_\mathrm{P}$. In addition, we show the ordering 
temperature $T_\mathrm{C}$ due to dipole interaction by neglecting the 
exchange coupling between islands. Evidently, a distinct variation of 
$T_\mathrm{C}$ near $\Theta_\mathrm{P}$ is not obtained in this case. 

The nonequilibrium behavior as caused by the 
anisotropy $K$ differs strongly for coverages below and above 
$\Theta_\mathrm{P}$. Due to the slow relaxation of the irregular atomic 
structure for $\Theta < \Theta_\mathrm{P}$, a blocking temperature 
$T_\mathrm{b}(\Theta)$ is obtained which is an order of magnitude larger 
than $T_\mathrm{C}(\Theta)$ resulting from the dipole interaction. 
Evidently, $T_\mathrm{b}$ depends on the anisotropy $K$ and the MC time $t$.
On the other hand, for $\Theta>\Theta_\mathrm{P}$, the relaxation is 
accelerated by the exchange interaction. With increasing $\Theta$ the remanent 
magnetization reaches the equilibrium value within $t=1000$~MCS, hence
$T_\mathrm{b}(\Theta)$ merges into $T_\mathrm{C}(\Theta)$. 

Recently, a mean field theory (MFT) for the dipole-coupling induced 
magnetic ordering temperature has been performed, using a simplified 
growth model.\cite{poulopoulos}
A qualitatively similar behavior of $T_\mathrm{C}(\Theta)$ 
as compared to the present MC calculations was obtained, in particular 
the strong variation of  $T_\mathrm{C}$ near $\Theta_\mathrm{P}$ due to the 
exchange interaction between coagulated islands. 
Evidently, MFT yields much larger values for $T_\mathrm{C}(\Theta)$ 
for such low-dimensional systems especially for $\Theta<\Theta_\mathrm{P}$, 
since thermal fluctuations are neglected. 

In the following, we discuss our results in connection with
measurements on Co/Cu(001) ultrathin films. 
Although several model parameters are chosen in accordance with this system,
a full quantitative comparison cannot be drawn yet. The main reason is that
the observed intermixing of Co adatoms with Cu substrate atoms is not taken 
into account within our growth model due to the incomplete knowledge
of the resulting atomic morphology. The measured percolation threshold 
$\Theta_\mathrm{P}\approx1.7$~ML is much larger than the one as obtained 
with the growth parameters used by us. 
 
To investigate solely the effect of an enlarged $\Theta_\mathrm{P}$, 
we have performed additional simulations 
simply by taking into account magnetic islands with up to three atomic layers, 
yielding the observed  $\Theta_\mathrm{P}$. 
Then for a coverage $\Theta=1.6$~ML, the dipole coupling induces 
a ordering temperature $T_\mathrm{C}\approx50$~K.  
The corresponding blocking temperature for $K=0.01$~meV and $t=1000$~MCS 
is obtained to be $T_\mathrm{b}\approx150$~K. 
Hence, we find that for the assumed growth modes and magnetic parameters 
the  blocking temperatures are always markedly larger than the ordering 
temperatures. These temperatures are comparable with measured 
temperatures $T_\mathrm{C}\approx 150$~K
for coverages slightly below $\Theta_\mathrm{P}$.\cite{bovensiepen}
In addition, to draw a quantitative comparison with the Co/Cu(001) 
system  a fourfold symmetry of the in-plane anisotropy  
has to be taken into account. We expect that the above stated general 
behavior of the magnetic relaxation and ordering obtained with a 
uniaxial anisotropy will not be changed. 

\section{\label{conclusion}Conclusion}
In this study, we have calculated the
nonequilibrium and equilibrium remanent magnetization of growing ultrathin
films, using a cluster Monte Carlo method. An island-type nanostructure 
with a nonuniform distribution of island sizes, shapes, and locations
was investigated. Within a micromagnetic model (modified Ising model) 
the single-island magnetic anisotropy, the  
dipole coupling, and the exchange interaction between magnetic islands 
were taken into account. We have analysed the transition from dipole-coupled 
islands for film coverages below the percolation 
threshold $\Theta_\mathrm{P}$ towards a connected ferromagnetic film above
$\Theta_\mathrm{P}$ with increasing film coverage.

For coverages $\Theta<\Theta_\mathrm{P}$, the dipole interaction leads to 
an equilibrium net magnetization refering to a collectively ordered state.  
A small ordering temperature $T_\mathrm{C}$ of about 1 -- 10~K results 
for the assumed model parameters. 
$T_\mathrm{C}$ increases strongly near $\Theta_\mathrm{P}$
due to exchange interaction which aligns coagulated islands.   
On the other hand, the anisotropy induces a pronounced 
nonequilibrium remanent magnetization which may be visible in experiment 
even after long waiting times. The corresponding blocking 
temperature $T_\mathrm{b}$ is obtained to be of the order of 10 -- 100~K, 
which is always markedly larger than $T_\mathrm{C}$.  
Approaching $\Theta_\mathrm{P}$, the proportionality  
between blocking temperature and anisotropy is no longer valid 
due to relaxation via internal island cluster excitations.   
A nonequilibrium remanent magnetization due to anisotropy is visible also 
for coverages $\Theta\gtrsim\Theta_\mathrm{P}$ where the film is still 
very irregular. For smoother films at larger coverages the exchange 
interaction induces a fast magnetic relaxation towards equilibrium.

We have obtained these results with a cluster-spin-flip algorithm which 
takes into account coherent magnetic rotations of island clusters. 
This method leads to a very fast and more realistic magnetic relaxation 
towards equilibrium in the coverage range with an  irregular nanostructure. 
Our results cannot be achieved by conventional single-spin-flip
algorithms. The suggested CSF algorithm can be applied also to
other inhomogeneous spin systems such as diluted magnets and spin glasses. 

Several possible improvements of our micromagnetic model are pointed out. 
In this study, we have used Ising-like states $S_i^x=\pm1$. 
By applying continuously varying spins $\bm{S}_i$, noncollinear 
magnetic arrangements can be analysed,\cite{jensen02a} allowing one to also 
determine the effects of an external magnetic field for these 
strongly inhomogeneous films. In particular, the movement of magnetic 
domain walls can be investigated. 
Furthermore, various magnetic nanostructures like 
chains and stripes\cite{shen97b}
can be studied easily by a proper variation of the parameters of the 
Eden-type growth model. Anisotropies with, e.g., a four-fold in-plane 
symmetry will be considered. Finally, the relaxation laws and times 
of the remanent magnetization can be investigated 
for such thin film systems.\cite{shen97b}  

\begin{acknowledgments} 
This work was supported by the Deutsche Forschungsgemeinschaft, 
SFB 290, TP A1, and 450, TP C4. Discussions with Prof.\ K.\ Baberschke and 
Dr.\ J.\ Lindner are acknowledged. 
\end{acknowledgments}

\bibliography{biblio}

\begin{thebibliography}{30}
\expandafter\ifx\csname natexlab\endcsname\relax\def\natexlab#1{#1}\fi
\expandafter\ifx\csname bibnamefont\endcsname\relax
  \def\bibnamefont#1{#1}\fi
\expandafter\ifx\csname bibfnamefont\endcsname\relax
  \def\bibfnamefont#1{#1}\fi
\expandafter\ifx\csname citenamefont\endcsname\relax
  \def\citenamefont#1{#1}\fi
\expandafter\ifx\csname url\endcsname\relax
  \def\url#1{\texttt{#1}}\fi
\expandafter\ifx\csname urlprefix\endcsname\relax\def\urlprefix{URL }\fi
\providecommand{\bibinfo}[2]{#2}
\providecommand{\eprint}[2][]{\url{#2}}

\bibitem[{\citenamefont{Himpsel et~al.}(1998)\citenamefont{Himpsel, Ortega,
  Mankey, and Willis}}]{himpsel}
\bibinfo{author}{\bibfnamefont{F.~J.} \bibnamefont{Himpsel}},
  \bibinfo{author}{\bibfnamefont{J.~E.} \bibnamefont{Ortega}},
  \bibinfo{author}{\bibfnamefont{G.~J.} \bibnamefont{Mankey}},
  \bibnamefont{and} \bibinfo{author}{\bibfnamefont{R.~F.}
  \bibnamefont{Willis}}, \bibinfo{journal}{Adv. Phys.}
  \textbf{\bibinfo{volume}{47}}, \bibinfo{pages}{511} (\bibinfo{year}{1998}).

\bibitem[{\citenamefont{Sun et~al.}(2000)\citenamefont{Sun, Murray, Weller,
  Folks, and Moser}}]{sun}
\bibinfo{author}{\bibfnamefont{S.}~\bibnamefont{Sun}},
  \bibinfo{author}{\bibfnamefont{C.~B.} \bibnamefont{Murray}},
  \bibinfo{author}{\bibfnamefont{D.}~\bibnamefont{Weller}},
  \bibinfo{author}{\bibfnamefont{L.}~\bibnamefont{Folks}}, \bibnamefont{and}
  \bibinfo{author}{\bibfnamefont{A.}~\bibnamefont{Moser}},
  \bibinfo{journal}{Science} \textbf{\bibinfo{volume}{287}},
  \bibinfo{pages}{198} (\bibinfo{year}{2000}).

\bibitem[{\citenamefont{Shen et~al.}(1997)\citenamefont{Shen, Skomski, Klaua,
  Jenniches, Manoharan, and Kirschner}}]{shen97b}
\bibinfo{author}{\bibfnamefont{J.}~\bibnamefont{Shen}},
  \bibinfo{author}{\bibfnamefont{R.}~\bibnamefont{Skomski}},
  \bibinfo{author}{\bibfnamefont{M.}~\bibnamefont{Klaua}},
  \bibinfo{author}{\bibfnamefont{H.}~\bibnamefont{Jenniches}},
  \bibinfo{author}{\bibfnamefont{S.~S.} \bibnamefont{Manoharan}},
  \bibnamefont{and}
  \bibinfo{author}{\bibfnamefont{J.}~\bibnamefont{Kirschner}},
  \bibinfo{journal}{Phys. Rev. B} \textbf{\bibinfo{volume}{56}},
  \bibinfo{pages}{2340} (\bibinfo{year}{1997}).

\bibitem[{\citenamefont{Pietzsch et~al.}(2000)\citenamefont{Pietzsch, Kubetzka,
  Bode, and Wiesendanger}}]{bode00}
\bibinfo{author}{\bibfnamefont{O.}~\bibnamefont{Pietzsch}},
  \bibinfo{author}{\bibfnamefont{A.}~\bibnamefont{Kubetzka}},
  \bibinfo{author}{\bibfnamefont{M.}~\bibnamefont{Bode}}, \bibnamefont{and}
  \bibinfo{author}{\bibfnamefont{R.}~\bibnamefont{Wiesendanger}},
  \bibinfo{journal}{Phys. Rev. Lett.} \textbf{\bibinfo{volume}{84}},
  \bibinfo{pages}{5212} (\bibinfo{year}{2000}).

\bibitem[{\citenamefont{Bovensiepen et~al.}(1999)\citenamefont{Bovensiepen,
  Poulopoulos, Platow, Farle, and Baberschke}}]{bovensiepen}
\bibinfo{author}{\bibfnamefont{U.}~\bibnamefont{Bovensiepen}},
  \bibinfo{author}{\bibfnamefont{P.}~\bibnamefont{Poulopoulos}},
  \bibinfo{author}{\bibfnamefont{W.}~\bibnamefont{Platow}},
  \bibinfo{author}{\bibfnamefont{M.}~\bibnamefont{Farle}}, \bibnamefont{and}
  \bibinfo{author}{\bibfnamefont{K.}~\bibnamefont{Baberschke}},
  \bibinfo{journal}{J. Magn. Magn. Mater.} \textbf{\bibinfo{volume}{192}},
  \bibinfo{pages}{L386} (\bibinfo{year}{1999}).

\bibitem[{\citenamefont{Poulopoulos et~al.}(2002)\citenamefont{Poulopoulos,
  Jensen, Ney, Lindner, and Baberschke}}]{poulopoulos}
\bibinfo{author}{\bibfnamefont{P.}~\bibnamefont{Poulopoulos}},
  \bibinfo{author}{\bibfnamefont{P.~J.} \bibnamefont{Jensen}},
  \bibinfo{author}{\bibfnamefont{A.}~\bibnamefont{Ney}},
  \bibinfo{author}{\bibfnamefont{J.}~\bibnamefont{Lindner}}, \bibnamefont{and}
  \bibinfo{author}{\bibfnamefont{K.}~\bibnamefont{Baberschke}},
  \bibinfo{journal}{Phys. Rev. B} \textbf{\bibinfo{volume}{65}},
  \bibinfo{pages}{064431} (\bibinfo{year}{2002}), \bibinfo{note}{and references
  therein}.

\bibitem[{\citenamefont{N{\'e}el}(1949)}]{neel}
\bibinfo{author}{\bibfnamefont{L.}~\bibnamefont{N{\'e}el}},
  \bibinfo{journal}{Ann. G{\'e}ophys. (C.N.R.S.)} \textbf{\bibinfo{volume}{5}},
  \bibinfo{pages}{99} (\bibinfo{year}{1949}).

\bibitem[{\citenamefont{Brown}(1959)}]{brown1}
\bibinfo{author}{\bibfnamefont{W.~F.} \bibnamefont{Brown}, \bibfnamefont{Jr.}},
  \bibinfo{journal}{J. Appl. Phys.} \textbf{\bibinfo{volume}{30}},
  \bibinfo{pages}{130S} (\bibinfo{year}{1959}).

\bibitem[{\citenamefont{Cowburn et~al.}(1999)\citenamefont{Cowburn, Adeyeye,
  and Welland}}]{cowburn}
\bibinfo{author}{\bibfnamefont{R.~P.} \bibnamefont{Cowburn}},
  \bibinfo{author}{\bibfnamefont{A.~O.} \bibnamefont{Adeyeye}},
  \bibnamefont{and} \bibinfo{author}{\bibfnamefont{M.~E.}
  \bibnamefont{Welland}}, \bibinfo{journal}{New J. Phys.}
  \textbf{\bibinfo{volume}{1}}, \bibinfo{pages}{16.1} (\bibinfo{year}{1999}).

\bibitem[{\citenamefont{Dormann et~al.}(1999)\citenamefont{Dormann, Fiorani,
  and Tronc}}]{dormann}
\bibinfo{author}{\bibfnamefont{J.~L.} \bibnamefont{Dormann}},
  \bibinfo{author}{\bibfnamefont{D.}~\bibnamefont{Fiorani}}, \bibnamefont{and}
  \bibinfo{author}{\bibfnamefont{E.}~\bibnamefont{Tronc}}, \bibinfo{journal}{J.
  Magn. Magn. Mater.} \textbf{\bibinfo{volume}{202}}, \bibinfo{pages}{251}
  (\bibinfo{year}{1999}), \bibinfo{note}{and references therein}.

\bibitem[{\citenamefont{Hansen and M{\o}rup}(1998)}]{morup}
\bibinfo{author}{\bibfnamefont{M.~F.} \bibnamefont{Hansen}} \bibnamefont{and}
  \bibinfo{author}{\bibfnamefont{S.}~\bibnamefont{M{\o}rup}},
  \bibinfo{journal}{J. Magn. Magn. Mater.} \textbf{\bibinfo{volume}{184}},
  \bibinfo{pages}{262} (\bibinfo{year}{1998}), \bibinfo{note}{and references
  therein}.

\bibitem[{\citenamefont{Mamiya et~al.}(1998)\citenamefont{Mamiya, Nakatani, and
  Furubayashi}}]{mamiya}
\bibinfo{author}{\bibfnamefont{H.}~\bibnamefont{Mamiya}},
  \bibinfo{author}{\bibfnamefont{I.}~\bibnamefont{Nakatani}}, \bibnamefont{and}
  \bibinfo{author}{\bibfnamefont{T.}~\bibnamefont{Furubayashi}},
  \bibinfo{journal}{Phys. Rev. Lett.} \textbf{\bibinfo{volume}{80}},
  \bibinfo{pages}{177} (\bibinfo{year}{1998}).

\bibitem[{\citenamefont{Poddar et~al.}(2002)\citenamefont{Poddar, Telem-Shafir,
  Fried, and Markovich}}]{poddar}
\bibinfo{author}{\bibfnamefont{P.}~\bibnamefont{Poddar}},
  \bibinfo{author}{\bibfnamefont{T.}~\bibnamefont{Telem-Shafir}},
  \bibinfo{author}{\bibfnamefont{T.}~\bibnamefont{Fried}}, \bibnamefont{and}
  \bibinfo{author}{\bibfnamefont{G.}~\bibnamefont{Markovich}},
  \bibinfo{journal}{Phys. Rev. B} \textbf{\bibinfo{volume}{66}},
  \bibinfo{pages}{060403(R)} (\bibinfo{year}{2002}).

\bibitem[{\citenamefont{Jonsson et~al.}(1998)\citenamefont{Jonsson, Svedlindh,
  and Hansen}}]{jonsson}
\bibinfo{author}{\bibfnamefont{T.}~\bibnamefont{Jonsson}},
  \bibinfo{author}{\bibfnamefont{P.}~\bibnamefont{Svedlindh}},
  \bibnamefont{and} \bibinfo{author}{\bibfnamefont{M.~F.}
  \bibnamefont{Hansen}}, \bibinfo{journal}{Phys. Rev. Lett.}
  \textbf{\bibinfo{volume}{81}}, \bibinfo{pages}{3976} (\bibinfo{year}{1998}).

\bibitem[{\citenamefont{Lottis and White}(1991)}]{lottis}
\bibinfo{author}{\bibfnamefont{D.~K.} \bibnamefont{Lottis}} \bibnamefont{and}
  \bibinfo{author}{\bibfnamefont{R.~M.} \bibnamefont{White}},
  \bibinfo{journal}{Phys. Rev. Lett.} \textbf{\bibinfo{volume}{67}},
  \bibinfo{pages}{362} (\bibinfo{year}{1991}).

\bibitem[{\citenamefont{Hilo et~al.}(1994)\citenamefont{Hilo, O'Grady, and
  Chantrell}}]{chantrell}
\bibinfo{author}{\bibfnamefont{M.~E.} \bibnamefont{Hilo}},
  \bibinfo{author}{\bibfnamefont{K.~O.} \bibnamefont{O'Grady}},
  \bibnamefont{and} \bibinfo{author}{\bibfnamefont{R.~W.}
  \bibnamefont{Chantrell}}, \bibinfo{journal}{J. Appl. Phys.}
  \textbf{\bibinfo{volume}{76}}, \bibinfo{pages}{6811} (\bibinfo{year}{1994}).

\bibitem[{\citenamefont{Brinzanik et~al.}(2002)\citenamefont{Brinzanik, Jensen,
  and Bennemann}}]{dompub}
\bibinfo{author}{\bibfnamefont{R.}~\bibnamefont{Brinzanik}},
  \bibinfo{author}{\bibfnamefont{P.~J.} \bibnamefont{Jensen}},
  \bibnamefont{and} \bibinfo{author}{\bibfnamefont{K.~H.}
  \bibnamefont{Bennemann}}, \bibinfo{journal}{J. Magn. Magn. Mater.}
  \textbf{\bibinfo{volume}{238}}, \bibinfo{pages}{258} (\bibinfo{year}{2002}).

\bibitem[{\citenamefont{Brinzanik}(2003)}]{phd}
\bibinfo{author}{\bibfnamefont{R.}~\bibnamefont{Brinzanik}},
  \bibinfo{type}{{Ph.D. thesis}}, \bibinfo{school}{Freie Universit{{\"a}}t
  Berlin} (\bibinfo{year}{2003}), \bibinfo{note}{electronic version at URL
  http://www.diss.fu-berlin.de/2003/253/indexe.html}.

\bibitem[{\citenamefont{Stoner and Wohlfarth}(1947)}]{stonerw47}
\bibinfo{author}{\bibfnamefont{E.~C.} \bibnamefont{Stoner}} \bibnamefont{and}
  \bibinfo{author}{\bibfnamefont{E.~P.} \bibnamefont{Wohlfarth}},
  \bibinfo{journal}{Nature (London)} \textbf{\bibinfo{volume}{160}},
  \bibinfo{pages}{650} (\bibinfo{year}{1947}).

\bibitem[{\citenamefont{Wolff}(1989)}]{wolff}
\bibinfo{author}{\bibfnamefont{U.}~\bibnamefont{Wolff}},
  \bibinfo{journal}{Phys. Rev. Lett.} \textbf{\bibinfo{volume}{62}},
  \bibinfo{pages}{361} (\bibinfo{year}{1989}).

\bibitem[{\citenamefont{Davis et~al.}(1999)\citenamefont{Davis, McCausland,
  McGahagan, and Widom}}]{davis}
\bibinfo{author}{\bibfnamefont{S.~W.} \bibnamefont{Davis}},
  \bibinfo{author}{\bibfnamefont{W.}~\bibnamefont{McCausland}},
  \bibinfo{author}{\bibfnamefont{H.~C.} \bibnamefont{McGahagan}},
  \bibnamefont{and} \bibinfo{author}{\bibfnamefont{M.}~\bibnamefont{Widom}},
  \bibinfo{journal}{Phys. Rev. E} \textbf{\bibinfo{volume}{59}},
  \bibinfo{pages}{2424} (\bibinfo{year}{1999}).

\bibitem[{\citenamefont{Landau and Binder}(2000)}]{landaubinder}
\bibinfo{author}{\bibfnamefont{D.~P.} \bibnamefont{Landau}} \bibnamefont{and}
  \bibinfo{author}{\bibfnamefont{K.}~\bibnamefont{Binder}},
  \emph{\bibinfo{title}{A Guide to Monte Carlo Simulations in Statistical
  Physics}} (\bibinfo{publisher}{Cambridge University Press},
  \bibinfo{address}{Cambridge}, \bibinfo{year}{2000}).

\bibitem[{\citenamefont{Jensen}(1997)}]{jensen97}
\bibinfo{author}{\bibfnamefont{P.~J.} \bibnamefont{Jensen}},
  \bibinfo{journal}{Ann. Phys. (Leipzig)} \textbf{\bibinfo{volume}{6}},
  \bibinfo{pages}{317} (\bibinfo{year}{1997}).

\bibitem[{\citenamefont{Srivastava et~al.}(1998)\citenamefont{Srivastava,
  Wilhelm, Ney, Farle, Wende, Haack, Ceballos, and Baberschke}}]{srivastava}
\bibinfo{author}{\bibfnamefont{P.}~\bibnamefont{Srivastava}},
  \bibinfo{author}{\bibfnamefont{F.}~\bibnamefont{Wilhelm}},
  \bibinfo{author}{\bibfnamefont{A.}~\bibnamefont{Ney}},
  \bibinfo{author}{\bibfnamefont{M.}~\bibnamefont{Farle}},
  \bibinfo{author}{\bibfnamefont{H.}~\bibnamefont{Wende}},
  \bibinfo{author}{\bibfnamefont{N.}~\bibnamefont{Haack}},
  \bibinfo{author}{\bibfnamefont{G.}~\bibnamefont{Ceballos}}, \bibnamefont{and}
  \bibinfo{author}{\bibfnamefont{K.}~\bibnamefont{Baberschke}},
  \bibinfo{journal}{Phys. Rev. B} \textbf{\bibinfo{volume}{58}},
  \bibinfo{pages}{5701} (\bibinfo{year}{1998}).

\bibitem[{\citenamefont{Stauffer}(1985)}]{stauffer}
\bibinfo{author}{\bibfnamefont{D.}~\bibnamefont{Stauffer}},
  \emph{\bibinfo{title}{Introduction to Percolation Theory}}
  (\bibinfo{publisher}{Taylor \& Francis}, \bibinfo{address}{London and
  Philadelphia}, \bibinfo{year}{1985}).

\bibitem[{\citenamefont{Hoshen and Kopelman}(1976)}]{hoshen}
\bibinfo{author}{\bibfnamefont{J.}~\bibnamefont{Hoshen}} \bibnamefont{and}
  \bibinfo{author}{\bibfnamefont{R.}~\bibnamefont{Kopelman}},
  \bibinfo{journal}{Phys. Rev. B} \textbf{\bibinfo{volume}{14}},
  \bibinfo{pages}{3438} (\bibinfo{year}{1976}).

\bibitem[{\citenamefont{De'Bell et~al.}(2000)\citenamefont{De'Bell, MacIsaac,
  and Whitehead}}]{macisaac00}
\bibinfo{author}{\bibfnamefont{K.}~\bibnamefont{De'Bell}},
  \bibinfo{author}{\bibfnamefont{A.~B.} \bibnamefont{MacIsaac}},
  \bibnamefont{and} \bibinfo{author}{\bibfnamefont{J.~P.}
  \bibnamefont{Whitehead}}, \bibinfo{journal}{Rev. Mod. Phys.}
  \textbf{\bibinfo{volume}{72}}, \bibinfo{pages}{225} (\bibinfo{year}{2000}).

\bibitem[{\citenamefont{Jensen and Pastor}(2002)}]{jensen02a}
\bibinfo{author}{\bibfnamefont{P.~J.} \bibnamefont{Jensen}} \bibnamefont{and}
  \bibinfo{author}{\bibfnamefont{G.~M.} \bibnamefont{Pastor}},
  \bibinfo{journal}{Phys. Stat. Sol. (a)} \textbf{\bibinfo{volume}{189}},
  \bibinfo{pages}{527} (\bibinfo{year}{2002}).

\bibitem[{\citenamefont{Jensen and Pastor}(2003)}]{jensen02b}
\bibinfo{author}{\bibfnamefont{P.~J.} \bibnamefont{Jensen}} \bibnamefont{and}
  \bibinfo{author}{\bibfnamefont{G.~M.} \bibnamefont{Pastor}},
  \bibinfo{journal}{New J. Phys.} \textbf{\bibinfo{volume}{5}},
  \bibinfo{pages}{68} (\bibinfo{year}{2003}).

\bibitem[{\citenamefont{Heinrich et~al.}(1991)\citenamefont{Heinrich, Cochran,
  Kowalewski, Kirschner, Celinski, Arrot, and Myrtle}}]{heinrich}
\bibinfo{author}{\bibfnamefont{B.}~\bibnamefont{Heinrich}},
  \bibinfo{author}{\bibfnamefont{J.~F.} \bibnamefont{Cochran}},
  \bibinfo{author}{\bibfnamefont{M.}~\bibnamefont{Kowalewski}},
  \bibinfo{author}{\bibfnamefont{J.}~\bibnamefont{Kirschner}},
  \bibinfo{author}{\bibfnamefont{Z.}~\bibnamefont{Celinski}},
  \bibinfo{author}{\bibfnamefont{A.~S.} \bibnamefont{Arrot}}, \bibnamefont{and}
  \bibinfo{author}{\bibfnamefont{K.}~\bibnamefont{Myrtle}},
  \bibinfo{journal}{Phys. Rev. B} \textbf{\bibinfo{volume}{44}},
  \bibinfo{pages}{9348} (\bibinfo{year}{1991}).

\end{thebibliography}

\newpage 

\end{document}